\documentclass[prb,showpacs,floatfix,amsfonts]{revtex4}
\usepackage{graphicx,graphics,color,epsfig} % Include figure files
\usepackage{bm}
\usepackage{amsmath}
\usepackage{amssymb}
\begin{document}

\title{Collective Excitations of Dirac Electrons in Graphene}

\author{Vadim Apalkov$^\dagger$, Xue-Feng Wang$^*$, and Tapash 
Chakraborty$^*$}
\affiliation{
$^\dagger$Department of Physics and Astronomy, Georgia State 
University, Atlanta, Georgia 30303, USA \\
$^*$Department of Physics and Astronomy, University of 
Manitoba, Winnipeg, MB R2T 2N2, Canada}

%\maketitle

%\begin{history}
%\received{Day Month Year}
%\revised{Day Month Year}
%\accepted{(Day Month Year)}
%\comby{(xxxxxxxxxx)}
%\end{history}

\begin{abstract}
Two-dimensional electrons in graphene are known to behave as
massless fermions with Dirac-Weyl type linear dispersion near
the Dirac crossing points. We have investigated the collective
excitations of this system in the presence or absence of an
external magnetic field. Unlike in the conventional two-dimensional
electron system, the $\nu=\frac1m$ fractional quantum Hall state
in graphene was found to be most stable in the $n=1$ Landau level.
In the zero field case, but in the presence of the spin-orbit
interaction, an undamped plasmon mode was found to exist in the
gap of the single-particle continuum.
\end{abstract}

%\keywords{graphene; quantum Hall effect; plasmons.}
\maketitle

\section{Introduction}
In recent experimental work, it has been possible to extract an
atomically thin, two-dimensional (2D) sheet of graphite -- graphene,
by micromechanical cleavage\cite{graphene_2D}. A two-dimensional
electron system in graphene exhibits many remarkable properties.
In the band structure calculations where electrons are treated
as hopping on a hexagonal lattice\cite{wallace}, one finds a
unique {\it linear} (relativistic type) energy dispersion near
the corners of the first Brillouin zone where the conduction
and valence bands meet. As a consequence, the low-energy excitations
follow the Dirac-Weyl equations for massless relativistic
particles\cite{ando}. In an external magnetic field, the spectrum
develops into Landau levels, each of which approximately fourfold
degenerate\cite{mcclure,landau_levels}. Recent discovery of the
quantum Hall effect in graphene\cite{qhe_expt,high_field,qhe_theory}
has resulted in intense activities\cite{phys_today} to unravel
the electronic properties of graphene that are distinctly different
from the conventional (or, as popularly called the `non-relativistic')
2D electron systems in semiconductor structures. In this paper, we
report on our investigation of the collective excitations of the 2D
electron gas in graphene in the presence or absence of an external
magnetic field.

Graphene has a honeycomb lattice structure of $sp^2$ carbon atoms,
with two atoms, A and B per unit cell (i.e., a two-dimensional
triangular Bravais lattice with a basis of two atoms). Each atom is
tied with its three nearest neighbors via strong $\sigma$ bonds
that are in the same plane with angles of 120$^\circ$. The $\pi$
orbit ($2p_z$) of each atom is perpendicular to the plane and
overlaps with the $\pi$ orbitals of the neighboring atoms that
results in the delocalized $\pi$ and $\pi^\ast$ bands. There is
only one electron in each $\pi$ orbit and the Fermi energy is
located between the $\pi$ and $\pi^\ast$ bands. The separation
distance between the nearest neighbor atoms is $a_{cc}=0.14$ nm
while the lattice constant is $a=\sqrt3a_{cc}=0.246$ nm. The
dynamics of electrons in graphene is described by a nearest-neighbor
tight-binding model\cite{wallace} that describes the hopping of
electrons between the $2p_z$ carbon orbitals. The first Brillouin
zone is hexagonal and at two of its inequivalent corners (the K
and K$^\prime$ points) the conduction and valence bands meet.
Graphene is often described as a two valley (K and K$^\prime$)
zero-gap semiconductor. The two valleys correspond to two chiralities
of the Weyl-Dirac fermions\cite{chiral}. Near these two points (the
so-called Dirac points), the electrons have a relativistic-like
dispersion relation, $\varepsilon_k=\pm\hbar v \vert {\bf k}\vert$
and obey the Dirac-Weyl equations for massless fermions. At the
vanishing gate voltage, the system is half-filled and the Fermi
level lies at the Dirac points.

\section{Collective modes}
In the following sections, we describe results of our work on the
collective excitations of electrons in graphene. First, we
discuss the case of the collective modes in the presence of a
strong perpendicular magnetic field. In particular, we discuss
the properties of the fractional quantum Hall states that reflect
the nature of electron correlations in the system in the presence
of a strong magnetic field. We found that the linear dispersion
of electrons discussed above, leads to a noticeable change in the
behavior than what is expected in a conventional two-dimensional
electron system. In the last section, we discuss the zero-field
case using the random-phase approximation and explore the
properties of plasmons in such a system.

\subsection{In a magnetic field}
An external magnetic field has a significant influence on the
energy spectrum of the 2D electron system in graphene. Details
of the single-electron case has been widely reported in the
literature\cite{ando,landau_levels}.

\subsubsection{Landau levels}
In the continuum limit the electron wave function in graphene
is a 8-component spinor, $\Psi_{s,k,\alpha}$, where $s=\pm 1/2$
is the spin index, $k = K, K^\prime$ is the valley index, and
$\alpha = A, B$ is the sublattice index. Without the spin-orbit
interaction\cite{kane05,wang06,sini} the spin degrees of freedom
becomes uncoupled from the spatial motion and the Hamiltonian of
an electron can be  described by two $4\times 4$ matrices for
each component of the electron spin. If we introduce the
four-component spinor as $(\Psi_{s,K,A},\Psi_{s,K,B},\Psi_{s,
K^\prime,A},\Psi_{s,K^\prime,B})$ then in the presence of a
magnetic field perpendicular to the graphene plane the Hamiltonian
matrix has the form
\begin{equation}
{\cal H} = v \left(
\begin{array}{cccc}
    0 & \pi_x - i \pi_y & 0 & 0   \\
    \pi_x + i \pi_y & 0 & 0 & 0 \\
    0 & 0& 0& \pi_x - i \pi_y   \\
    0 & 0&  \pi_x + i \pi_y & 0
\end{array}
\right) ,
\label{H}
\end{equation}
where ${\bm \pi}$=${\bm p} + e{\bm A}/c$, $\bm p$ is the
two-dimensional momentum, $\bm A$ is the vector potential,
and $v$ is the velocity of electrons in graphene. It is easy
to see from the Hamiltonian matrix (\ref{H}) that the valley
index in conserved. The conservation of the valley index is
easily violated in the graphene systems with a short-range
scattering impurity potential or in the many-body systems with
inter-electron interactions. In both cases the scattering of
the electron either by an impurity or by another electron introduces
the umklapp process and a change of the electron valley index.

Just as for the non-relativistic system the application of a
perpendicular magnetic field to the graphene layer results in the
Landau quantization. Due to the relativistic nature of electrons
in graphene the energy spectrum of the Landau levels has a unique
form, namely, the energy of the $n$th Landau level is
$E_n = sgn (n) \sqrt{ 2 e \hbar v^2 |n| B}$. This is different from
the non-relativistic electrons, where the Landau level spectrum is of
the harmonic oscillator type, i.e. equidistant, $E_n\propto n$.

In the ideal graphene system the eigenfunctions of the single-electron
Hamiltonian (\ref{H}) are specified by the Landau index, $n=0,\pm 1,
\pm 2, \ldots $ and an intra-Landau index $m$, which depends on the
gauge. Each Landau level is fourfold degenerate due to the spin and valley
degrees of freedom. The corresponding wave functions for an electron
in the valleys $K$ and $K^\prime$ are described by the vectors
\begin{equation}
\Psi_{K,n} = C_n
\left( \begin{array}{c}
 {\rm sgn} (n) i^{|n|-1} \phi_{|n|-1} \\
    i^{|n|} \phi _{|n|} \\
    0  \\
    0
\end{array}
 \right),
\label{f1}
\end{equation}
\begin{equation}
\Psi_{K^\prime,n} = C_n
\left( \begin{array}{c}
   0 \\
   0\\
    i^{|n|} \phi _{|n|} \\
 {\rm sgn} (n) i^{|n|-1} \phi_{|n|-1}
\end{array}
 \right),
\label{f2}
\end{equation}
where $C_n = 1 $ for $n=0$ and $C_n = 1/\sqrt{2}$ for $n\neq 0$.
The two non-zero terms in $\Psi_{K,n}\, (\Psi_{K^\prime,n})$
correspond to occupation of the sublattice A (the upper term)
and the sublattice B (the lower term). Here $\phi _n$ is the Landau
wave function for a particle with the non-relativistic parabolic
dispersion relation in the $n$-th Landau level. From
Eqs.~(\ref{f1})--(\ref{f2}) it is clear that a specific feature of
the relativistic dispersion law is the ``mixture'' of the
non-relativistic Landau levels. This mixture is present only
for $n\neq 0$. For $n=0$ the electron in the valley $K$ or
$K^\prime$  occupies only the sublattice $A$ or $B$, respectively.
For higher Landau levels the electron in each valley occupies both
sublattices, $A$ and $B$. The wave functions in the sublattices
$A$ and $B$ are the wave functions of the non-relativistic electrons
with different Landau level indices, i.e. the relativistic wave
function is the mixture of non-relativistic wave functions of
different Landau levels. As we shall see below, this property of
the relativistic electrons strongly modifies the inter-electron
interaction within a single relativistic Landau level.

\subsubsection{Inter-electron interaction}
In what follows, we shall consider only the partially occupied Landau
levels with fractional filling factors. In this case the ground state
of the system and the excitation spectrum are fully determined by the
inter-electron interactions, which are completely described by the
Haldane pseudopotentials\cite{haldane} $V_m$. Haldane pseudopotentials
are the energies of two electrons with relative angular momentum $m$.
The pseudopotentials for the $n$-th Landau level can be presented
as\cite{haldane}
\begin{equation}
V_m^{(n)} = \int _0^{\infty } \frac{dq}{2\pi} q V(q)
\left[F_n(q) \right]^2 L_m (q^2)
 e^{-q^2},
\label{Vm}
\end{equation}
where $L_m(x)$ are the Laguerre polynomials, $V(q) = 2\pi e^2/(\kappa l q)$
is the Coulomb interaction in the momentum space, $\kappa$ is the
dielectric constant, and $F_n(q)$ is the form factor corresponding to
the $n$-th Landau level. The main difference between the relativistic
and the non-relativistic electrons is in the expression for the form
factor, $F_n(q)$. For relativistic electrons the form factor is
given by the equations\cite{nomura,goerbig06}
\begin{eqnarray}
& & F_0(q) = L_0\left( \frac{q^2}2 \right)  \label{f0}  \\
& & F_{n\neq 0}(q) = \frac12 \left[ L_n \left(\frac{q^2}2 \right)
 + L_{n-1} \left( \frac{q^2}2 \right)    \right],
\label{fn}
\end{eqnarray}
while for the non-relativistic particles the form factors in
Eq.~(\ref{Vm}) are
$$F_n (q) = L_n\left( q^2/2\right).\label{form_non_rev}$$
Comparing these non-relativistic form factors [Eq.~(\ref{form_non_rev})]
with Eqs.~(\ref{f0})-(\ref{fn}), we see that the inter-electron
interactions for the relativistic and the non-relativistic electrons
are the same for $n=0$ and different for $n>0$.

\begin{figure}
\begin{center}\includegraphics[width=9cm]{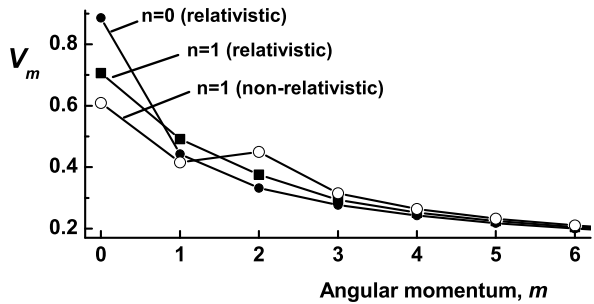}\end{center}
\vspace*{-1cm}
\caption{Pseudopotentials calculated from Eq.~(\ref{Vm}) are shown as
a function of the relative angular momentum  for relativistic
and non-relativistic 2D electrons for the
first two Landau levels. The energy is measured in units of $\epsilon _C$.
}
\label{figone}
\end{figure}

In Fig.~1 the pseudopotentials calculated from Eq.~(\ref{Vm}) for the
relativistic and the non-relativistic cases are shown. For $n=0$ the
non-relativistic and the relativistic pseudopotentials are the same.
The main feature of the pseudopotentials in the zeroth Landau level,
$V_m^{(0)}$, is their monotonic decrease with increasing relative
angular momentum, $m$.  Comparing $V_m^{(0)}$ to the non-relativistic
$V_m^{(1)}$ we can clearly see that (i) in higher Landau levels the
pseudopotentials become a non-monotonic function of $m$ and (ii)
the interaction strength is suppressed at $m=0$ and 1, i.e.
$V_m^{(1)}<V_m^{(0)}$, and enhanced at $m>1$, i.e. $V_m^{(1)}>V_m^{(0)}$.
From this behavior we can make the predictions about the stability and
excitation gaps of the FQHE. For example, for the $\nu = 1/3$-FQHE
the main parameter which determines the formation the incompressible
liquid is the ratio of the pseudopotentials at $m=3$ and $m=1$,
i.e., $V_3/V_1$. The smaller the ratio the larger the excitation gaps,
and consequently the more stable are the FQHE states. Comparing the
pseudopotentials of the non-relativistic electrons at $n=1$ and
$n=0$ we can conclude that the FQHE gaps in the first Landau level,
$n=1$ should be smaller then the corresponding gaps in the zeroth
Landau level, $n=0$.

The behavior of the pseudopotentials for the relativistic
electrons in the higher Landau levels is similar to that of the
relativistic ones. The exception is the properties of the
pseudopotential in the first Landau level, $n=1$. We notice in
Fig.~1 that the pseudopotentials in the first Landau level ($n=1$)
is larger than the corresponding pseudopotentials in the zeroth
Landau level, $n=0$, for all values of the relative angular
momentum, i.e. $V_m^{(1)}>V_m^{(0)}$. The pseudopotential in the
$n=1$ Landau level is also a monotonic function of $m$. This is
different from the non-relativistic case. Another important
property of the pseudopotential of the relativistic system is that
the ratio $V_3/V_1$ is the smallest in the first Landau level,
$n=1$. In Fig.~2 we present the results for the pseudopotentials
of the relativistic system in the lowest Landau levels. From this
figure we can also see that the ratio $V_3/V_1$ is the smallest
for $n=1$. Based on this property of $V_m^{(n)}$ we can conclude
that the largest gap of the 1/3-FQHE state should be expected in
the {\em first} Landau level, $n=1$. Therefore the FQHE in the
relativistic system should be easier to observe in the first
Landau level, but not in the zeroth Landau level as in the case of
the non-relativistic electrons.

\begin{figure}
\begin{center}\includegraphics[width=9cm]{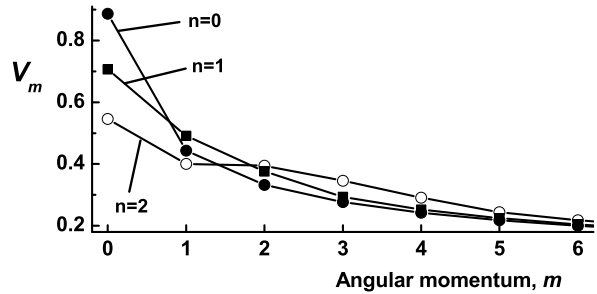}\end{center}
\vspace*{-1cm}
\caption{Pseudopotentials are shown as
a function of the relative angular momentum
for relativistic electrons in the first three lowest
Landau levels. The energy is measured in units of $\epsilon _C$.
}
\label{figtwo}
\end{figure}

Due to the presence of both the spin and the valley degeneracies the
graphene system becomes more complicated than the standard non-relativistic
electron system, where only the spin degree of freedom is present.
In a magnetic field the spin degeneracy can be lifted due to
the Zeeman energy, leaving the Landau levels of the graphene system
doubly degenerate due to the valley index.  Mathematically, the system
then becomes equivalent to a double layer non-relativistic system, where
the valley index is the layer index. Usually, in the double-layer
system  there is an asymmetry in the interaction Hamiltonian. This means
that the interaction strength between the electrons in the same layers
is different from that in different layers. This is due to a finite
separation between the layers. Comparing this property of a
double layer system to the graphene system we can say that the
graphene system is a double-layer system with zero separation
between the layers. So the system is completely SU(2)-pseudospin
symmetric, where the pseudospin is associated with the valley index.
But this is not entirely true. There is an asymmetry in the graphene
system as well. One of the types of asymmetry is related to
the lattice structure of graphene --- the interaction between the
electrons in the same sublattice is stronger than the
interaction between the electrons in different sublattices\cite{fisher06}.

To estimate the corrections due to the asymmetry related to the lattice
structure of graphene we assume that the electrons are localized
at the discrete points corresponding to the lattice structure of the
graphene. Then the interaction matrix element, $V_{i_1,j_1,i_2,j_2}$,
 between the single-electrons states, $\psi_i$,
 can be expressed in the following form
\begin{equation}
V_{i_1,j_1,i_2,j_2} = \sum_{{\bf r}_1} \sum_{{\bf r}_2}
\psi^{*}_{i_1}({\bf r}_1) \psi^{*}_{j_1}({\bf r}_2)
V({\bf r}_1-{\bf r}_2)
\psi_{j_2}({\bf r}_2) \psi_{i_2}({\bf r}_1)
\label{V_d1}
\end{equation}
where the index $i$ in the wavefunction $\psi_i$ indicates collectively
the Landau level, valley, sublattice, and intra-Landau level indices. The
sums in (\ref{V_d1}) run over the discrete positions of the electrons
corresponding to one of the sublattices of graphene. In the continuum
limit the sums should be replaced by the integral. Since two sublattices
of graphene are shifted by the vector ${\bf r}_0 = a(0,1/\sqrt3)$, in
the continuum limit the expression for the interaction matrix element
between the states belonging to different sublattices
should contain not the $V({\bf r}_1-{\bf r}_2)$ but
$V({\bf r}_1-{\bf r}_2-{\bf r}_0)$. This introduces the difference
in the interaction strength between the electrons in the same
sublattice and in different sublattices. In terms of the
pseudopotentials this means that if the pseudopotential is calculated
between the states of different sublattices then $V(q)$ in
the expression (\ref{Vm}) should contain an additional factor
$\exp({i{\bf q}\cdot{\bf r}_0})$ and the integral over $q$ should be
replaced by the 2D integral over $\bf{q}$.

In the zeroth Landau level, $n=0$, the electrons in the valley $K$
occupy sublattice $A$ only, while the electrons in the valley
$K^\prime$ occupy sublattice $B$. Then the pseudopotentials
corresponding to the interaction between the electrons belonging
to the same valley are given by the expression (\ref{Vm}) without
any modifications. The expression for the pseudopotentials
corresponding to the interaction between the electrons in the
different valleys should contain an additional factor and can be
written as
\begin{eqnarray}
V_{K,K^\prime,m}^{(0)} & = & \int \frac{d{\bf q}}{4\pi ^2}
V(q) e^{i{\bf q}\cdot{\bf r}_0/l} L_m (q^2)
e^{-q^2} \nonumber \\
 & = & V_{K,K,m}^{(0)} - \left( \frac{r_0}{l} \right)^2
\int \frac{q^2 d q}{4\pi }  V(q) L_m (q^2) e^{-q^2}
\label{VAm0}.
\end{eqnarray}
Since $r_0 = a/\sqrt{3}$ the asymmetric correction is proportional
to a small parameter $(a/l)^2$.

In the higher Landau levels the electrons in the valleys
$K$ and $K^\prime$ occupy both sublattices $A$ and $B$. This
results in the form factor of $\frac{1}{4} (L_n + L_{n-1})^2$ in
Eq.~(\ref{Vm}). The coefficients  $L_n$ and $L_{n-1}$ belong to different
sublattices. This can be schematically presented in the
following form: (i) for the intravalley interaction, we have
\begin{equation}
\frac{1}{4}\left[L_n (A) L_n(A) +L_{n-1} (B) L_{n-1}(B)
+2 L_n (A) L_{n-1}(B)  \right]
\end{equation}
and (ii) for the intervalley interaction,
\begin{equation}
\frac{1}{4}\left[L_n (A) L_n(B) +L_{n-1} (B) L_{n-1}(A)
+2 L_n (A) L_{n-1}(A) \right].
\end{equation}
Then following the same procedure as for the $n=0$ Landau level we
obtain the expressions for the pseudopotentials in the higher
Landau levels as
\begin{equation}
V_{K,K,m}^{(n)} = V_{K^\prime,K^\prime,m}^{(n)}  =
 \int \frac{d{\bf q}}{4\pi ^2}  V(q) L_m (q^2)  e^{-q^2}
\left[ F_n(q) + \frac12
L_n  L_{n-1} \left(e^{i{\bf q}\cdot{\bf r}_0/l}-1\right)\right],
\label{VAmn}
\end{equation}
and
\begin{equation}
V_{K,K^\prime,m}^{(n)} =
 \int \frac{d{\bf q}}{4\pi ^2}  V(q) L_m (q^2)  e^{-q^2}
\left[ F_n(q)
  e^{i{\bf q}\cdot{\bf r}_0/l} + \frac12
L_n L_{n-1}  \left(1-e^{i{\bf q}\cdot{\bf r}_0/l} \right) \right].
\label{VAmn2}
\end{equation}
It is convenient to rewrite the expressions
(\ref{VAmn}) and (\ref{VAmn2}) as
\begin{equation}
\tilde{V}_{K,K^\prime,m}^{(n)}= \tilde{V}_{K,K,m}^{(n)}
-\int \frac{d{\bf q}}{(4\pi)^2}V(q)L_m(q^2)e^{-q^2}
\left[L_n-L_{n-1}\right]^2\left(1-e^{i{\bf q}\cdot{\bf r}_0/l}\right).
\label{VAmn3}
\end{equation}
Similar to the zeroth Landau level the asymmetry correction in
Eq.~(\ref{VAmn3}) is proportional to a small parameter $(a/l)^2$.

Since the Coulomb interaction between the electrons in graphene
does not conserve the valley index there is an other mechanism for
violation of the SU(2) valley symmetry of the graphene system.
This mechanism is related to the inter-valley scattering. In the
lowest order in the small parameter $a/l$ the main scattering process
is the backscattering\cite{goerbig06}. During this process the
electron from the  $K$ valley is scattered into the $K^\prime$ valley,
while the electron from the  $K^\prime$ valley is scattered into the
$K$ valley. The interaction matrix elements corresponding to this
process are determined by the pseudopotentials (\ref{Vm}) where the
Coulomb interaction in the integral should be replaced by
$V({\bf q}+l\Delta{\bf K} )$. Here $\Delta{\bf K}={\bf K} -
{\bf K}^\prime = (2\pi /a) (-1/3, 1/\sqrt3)$. Therefore the
pseudopotentials describing the backscattering process are
given by the expression
\begin{equation}
V_{B,m}^{(n)} = \int \frac{d{\bf q}}{4\pi^2}
V({\bf q}+l\Delta{\bf K})
\left[F_n(q) \right]^2 L_m (q^2)e^{-q^2}.
\label{VBm}
\end{equation}
The leading order term in $V_{B,m}^{(n)}$ can be found by simply replacing
$V({\bf q}+l\Delta{\bf K})$ by $V(l\Delta{\bf K})\propto (a/l)$.

\subsubsection{FQHE in graphene}
With the pseudopotentials for Dirac electrons at hand, we now evaluate
numerically
the energy spectra of the many-electron states at the fractional fillings
of the Landau level. The calculations have been done in the spherical
geometry\cite{haldane} with the pseudopotentials given by Eq.~(\ref{Vm}).
In the spherical geometry the radius of the sphere $R$ is related to
$2S$ of magnetic fluxes through the sphere in units of the flux quanta
as $R = \sqrt{S} l$.  Here $2S$ is an integer number. The
single-electron states are characterized by the angular momentum $S$,
and its $z$ component, $S_z$. Therefore, at a given magnetic field,
i.e for a given flux $2S$, the number of available states in a sphere is
$(2S+1)$. Then for a given number of electrons $N$ the
parameter $S$ determines the filling factor of the Landau level. Due to
the spherical symmetry of the problem, the many-particle states are described
by the total angular momentum $L$ and its $z$ component, while the
energy depends only on $L$. The energy spectra of a many-particle
system is found by the standard procedure of calculating numerically
the lowest eigenvalues and eigenvectors of
the interaction Hamiltonian matrix\cite{fano}.

\begin{figure}
\begin{center}\includegraphics[width=9cm]{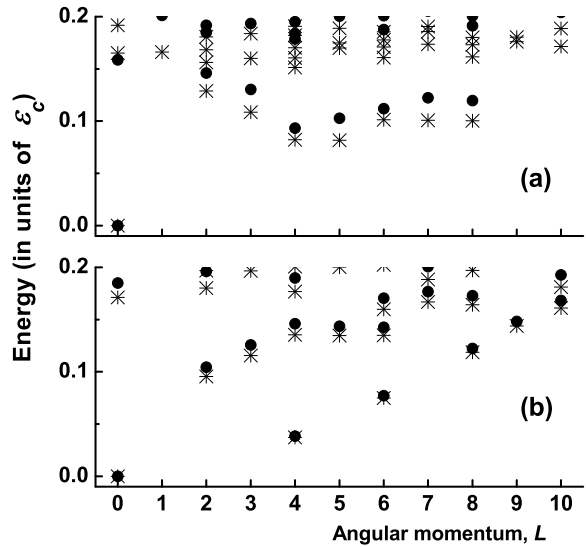}\end{center}
\vspace*{-1cm}
\caption{(a) The energy spectra of an eight-electron $\nu =1/3$-FQHE
system shown for different Landau levels:  $n=0$ (stars) and
$n=1$ (filled circles).  The flux quanta is $2S = 21$. (b)
Energy spectra of the six-electron $\nu =1/5$-FQHE
system is shown for different Landau levels:  $n=0$ (stars) and
$n=1$ (filled circles).  The flux quanta here is $2S = 25$.
}
\label{figthree}
\end{figure}

It was shown in the previous section that based on the analysis of
the Haldane pseudopotentials we can conclude that the FQHE states
in graphene should have the largest gaps in the $n=1$ Landau
level. Here we check this statement for $\nu =1/m$ incompressible
states. In the spherical geometry, such states are realized at $S
= (m/2)(N-1)$. The $1/m$ state in the higher $n$-th Landau level
is defined as a state corresponding to the $1/m$ filling factor of
the $n$-th Landau level, while all the lower energy Landau levels
are completely occupied. If the electron system is fully spin and
valley polarized then we should expect that the ground state to be
the Laughlin state\cite{laughlin,fqhe_book} which is separated
from the excited states by a finite gap.

In Fig.~3(a) the calculated energy spectra are shown for the $1/3$-FQHE
state and for different Landau levels.  Since the relativistic
pseudopotentials, $V_m^{(0)}$, for $n=0$ Landau level is similar to that of
non-relativistic case, the $1/3$ state and the corresponding energy gap
will be the same in both cases. The deviation from the non-relativistic
system occurs at the higher Landau levels. From  Fig.~3(a) we can
clearly see that the energy gap of the $1/3$-state at the $n=1$ Landau
level is enhanced when compared to that of the $n=0$ Landau level. At
higher Landau levels, i.e., for $n>1$, the excitation gaps are suppressed,
which means that at the $n=1$ Landau level the electron system in
graphene has the strongest interaction with the largest incompressible
gap \cite{gr-fqhe}. This is different from the non-relativistic case,
where the energy gap monotonically decreases with increasing Landau
level index\cite{fqhe_book}. At other filling factors of the type
$\nu = 1/m$ we should also expect the same increase of the energy gaps
at the $n=1$ Landau levels. The effect is however not as pronounced
as for $\nu=\frac13$ since at smaller filling factors the pseudopotentials
with a larger relative angular momentum, $m$, becomes important, for
which the difference between the pseudopotentials at $n=0$ and $n=1$
Landau levels is small [see Fig.~2]. This behavior is illustrated in
Fig.~3(b), where the results for the $\nu=1/5$ state are shown. We can
see that the difference between the excitation spectra at $n=0$ and
$n=1$ Landau levels is much smaller for this filling factor.
%The stronger interaction effects in the first Landau
%level, $n=1$, can be seen only at the higher excited states of the system.

\begin{figure}
\begin{center}\includegraphics[width=9cm]{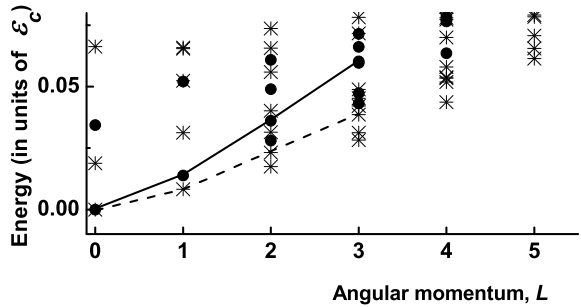}\end{center}
\vspace*{-1cm}
\caption{The energy spectra of a six-electron valley-unpolarized
$\nu =1/3$-FQHE system shown for different Landau levels: $n=0$ (stars)
and $n=1$ (dots). The flux quanta is $2S = 15$. The spin-wave excitations
are illustrated by solid ($n=1$) and dashed ($n=0$) lines.
}
\label{figfour}
\end{figure}

The results shown in Fig.~3 correspond to the fully spin and
valley polarized systems. This polarization is achieved at a high
magnetic field due to the Zeeman splitting and the valley
asymmetry. It is well known that even without the Zeeman energy
the ground state of the non-relativistic system at $\nu=1/m$ is
fully spin-polarized\cite{fqhe_book,fqhe_review} with the spin
equal to $S=N/2$, where $N$ is the number of particles. The
situation is the same in graphene: the ground state of the
$\nu=1/m$ liquid is fully spin and valley polarized. The results
shown Fig.~3 describe the polarized excitations in such a system.
Another type of neutral excitations of the incompressible liquid
is the spin or pseudo-spin (valley) excitations. To explore these
excitations we present the results of our calculations of the
energy spectra for the $\nu=1/3$ `unpolarized' system. Here we
assume that the system is fully spin-polarized but
valley-unpolarized. To characterize the states in such a system we
consider the valley index as a pseudospin, $\tau$. Results of
these calculations are shown in Fig.~4 for the zeroth and for the
first Landau level. The ground state is at zero angular momentum
and has a pseudospin equal to $\tau = N/2$, i.e. the
pseudospin-polarized ground state. One type of excitations in a
such system is the spin-waves. They are marked by the solid ($n=1$
Landau level) and dashed ($n=0$ Landau level) lines in Fig.~4.
Another type of excitations corresponds to the energy branch
formed by the lowest states at each angular momentum. The specific
feature of these states is that the angular momentum of the state
is related to the pseudospin value by the expression $(N/2)-\tau =
L$. The physical meaning of these states is the Bose condensation
of $L$ noninteracting pseudospin waves\cite{wojs}. Clearly, in all
the cases the energy scale at $n=1$ is larger than at $n=0$, which
again illustrates the stronger interaction effects at $n=1$.
However, this is not the general rule for the $n=1$ Landau level
since we have seen from Fig.~2 that the pseudopotential at the
zero relative momentum is stronger at the zeroth Landau level.
This pseudopotential becomes important only for the unpolarized
states when two electrons occupy the same spatial point, i.e. they
have zero relative angular momentum. For those states the
interaction effects can be stronger at the zeroth Landau level, as
is the case of the $\nu=\frac23$ incompressible
state\cite{gr-fqhe}. The excitation gap of the unpolarized
$\nu=\frac23$ state at the $n=0$ Landau level is larger than the
excitation gap at the $n=1$ Landau level\cite{gr-fqhe}.

In the above calculations we have also included the valley
asymmetry terms discussed in the previous section [see
Eqs.~(\ref{VAm0})-(\ref{VBm})]. These corrections result
into a very small shift of the energy levels up to magnetic field
of 50 Tesla. Indeed, these terms are of the order of $a/l$.
The magnetic length at $B=50$ T is about 3.6 nm, while the lattice
constant of graphene is $a = 0.246 $ nm. The parameter $a/l$
is then really small ($\sim$0.07).

Finally, from the results presented above we conclude that the
$1/m$-FQHE state in graphene is most stable at the $n=1$ Landau
level. The inter-electron interaction effects are therefore more
pronounced at the $n=1$ Landau level\cite{gr-fqhe}. This tendency
is just the opposite to that of the non-relativistic system, where
the excitation gap decreases monotonically with increasing Landau
level index. The enhancement of the interaction effects at the
higher Landau level index however depends on the filling factor.

\subsection{Zero magnetic field}
In the pseudospin space, the zero-magnetic-field Hamiltonian
of a spin-up electron with a wavevector around the $K$ point
is\cite{kane05,wang06,sini} ${\cal H} = v\bm{p}\cdot\bm{\sigma}+
\Delta_{\rm so}\sigma_z $ with $\bm{\sigma}=(\sigma_x,\sigma_y,
\sigma_z)$ the Pauli matrices and $\bm{p}$ the momentum operator.
Here $\Delta_{\rm so}$ is the strength of the spin-orbit interaction
(SOI). The eigenstates of the Schr\"{o}dinger equation $H\Psi=E\Psi$
are readily obtained as $\Psi_{\bm{k}}^\lambda(
\bm{r})=e^{i\bm{k}\cdot\bm{r}}{\small\left( \array{c}
1+\sin(\alpha_{\bm{k}}+\lambda\pi/2) \\
-e^{i\phi_{\bm{k}}}\cos(\alpha_{\bm{k}}+\lambda\pi/2)
\endarray
\right)}$ with energy $E_{\bm{k}}^\lambda=\lambda\sqrt{\Delta_{\rm
so}^2+\hbar^2v^2k^2}$ for $\lambda=+1$ denoting the conduction band and
$\lambda=-1$ the valance band. Here $\tan\phi_k=k_y/k_x, \tan\alpha_k
=\hbar v k/\Delta_{\rm so}$, and $k=\sqrt{k_x^2+k_y^2}$.

%Fig. 1
\begin{figure}[bt]
\centerline{\psfig{file=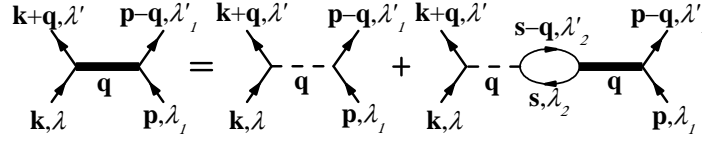,width=3.65in}}
\vspace*{8pt}
\caption{Diagrammatic illustration of the RPA dressed Coulomb
interaction. }
\label{fig:diagram}
\end{figure}

Using the techniques developed for the multicomponent
systems\cite{vint,wang1}, it is straightforward to show that
the RPA Coulomb interaction in the Fourier space $U(q,\omega)$
obeys the equation
\begin{equation}
U(q,\omega)=v_0+v_0\hat{\Pi}_0(q,\omega)U(q,\omega)
\end{equation}
with the electron-hole propagator
\begin{equation}
\hat{\Pi}_0(q,\omega) =4\sum_{\lambda, \lambda', \bm{k}}
|g_{\bm{k}}^{\lambda,\lambda'}(\bm{q})|^2
%\Pi_{\lambda, \lambda'}_{\bf k}({\bf q},\omega)
\frac{f[E^{\lambda'}_{\bm{k}+\bm{q}}]-f[E^\lambda_{\bm{k}}]}
{\omega+E^{\lambda'}_{\bm{k}+\bm{q}}-E^\lambda_{\bm{k}}+i\delta},
\label{propagator}
\end{equation}
as illustrated by the Feymann diagram in Fig.~\ref{fig:diagram}.
Here $v_0=e^2/(2\epsilon_0\epsilon_i q)$ is the two-dimensional
Coulomb interaction (in Fourier space) with the high-frequency
dielectric constant\cite{pere} $\epsilon_i=1$ and
$g_{\bm{k}}^{\lambda,\lambda'}(\bm{q})$ is the interaction vertex.

%Fig. 2
\begin{figure}[bt]
\centerline{\psfig{file=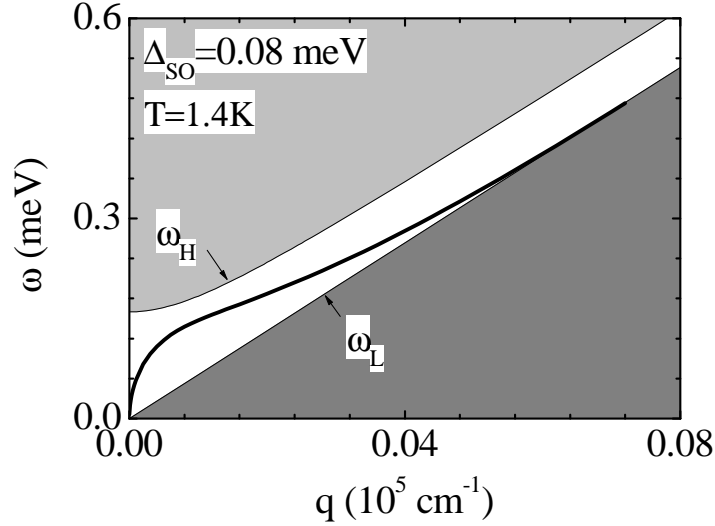,width=3.65in}}
\vspace*{8pt}
\caption{Plasmon spectrum (thick curves) of an electron gas in an
intrinsic graphene ($E_F=0$) at temperatures $T=1.42$ K with $\Delta_{
\rm so}=0.08$ meV. Intra- (dark shaded) and inter- (light shaded) band
single-particle continuums are also shown. $\omega_L$ and $\omega_H$
are the lower and upper borders separating the white (EHC gap) and
shaded areas respectively.}
\label{fig:plasmon}
\end{figure}

The factor four in Eq.~(\ref{propagator}) comes from the degenerate
two spins and two valleys at $K$ and $K^\prime$; the vertex factor reads
$|g^{\lambda,\lambda'}_{\bm{k}}(\bm{q})|^2=
[1+\lambda\lambda'\cos\alpha_{\bm{k}+\bm{q}}\cos\alpha_{\bm{k}}
+\lambda\lambda'\sin\alpha_{\bm{k}+\bm{q}}\sin\alpha_{\bm{k}} (k+q
\cos\theta)/|\bm{k}+\bm{q}|]/2$ with $\theta$ being the angle
between $\bm{k}$ and $\bm{q}$. Since the chiral property of the
system prohibits the intra-band backward scattering at $\bm{q}=2\bm{k}$
and the inter-band vertical transition at $\bm{q}=0$ under the
Coulomb interaction in the system, we have $|g^{\lambda,-\lambda}_{
\bm{k}}(0)|^2=|g^{\lambda,\lambda}_{\bm{k}} (2\bm{k})|^2=0$. The
collective excitation spectrum is obtained by finding the zeros of
the real part of the dielectric function $ \hat{\epsilon}(q,\omega)=1
-v_0(q)\hat{\Pi}_0(q,\omega) \label{dielsg}$.

In the presence of the SOI, an energy gap opens between
the conduction and valence bands and the semimetal electronic
system in graphene is converted into a narrow gap semiconductor
system. At the same time, a gap is opened between its intraband
single-particle continuum $\omega \leq \omega_L\equiv\hbar vq$
and its interband single-particle continuum $\omega \geq \omega_H
\equiv 2\sqrt{\Delta_{\rm so}^2+\hbar^2v^2q^2/4}$. However,
the system differs from a normal narrow gap semiconductor due to
its peculiar chiral property. In this paper, we have chosen
the magnitude of the SOI strength to be around $0.08 - 0.1$ meV
in graphene\cite{kane05,mele}. The result can be easily applied
to Dirac gases with different $\Delta_{\rm so}$ by scaling the
energy and wavevector in units of $\Delta_{\rm so}$ and
$k_{\rm so}=\Delta_{\rm so}/(\hbar v)$ respectively.

%Fig. 3
\begin{figure}[bt]
\centerline{\psfig{file=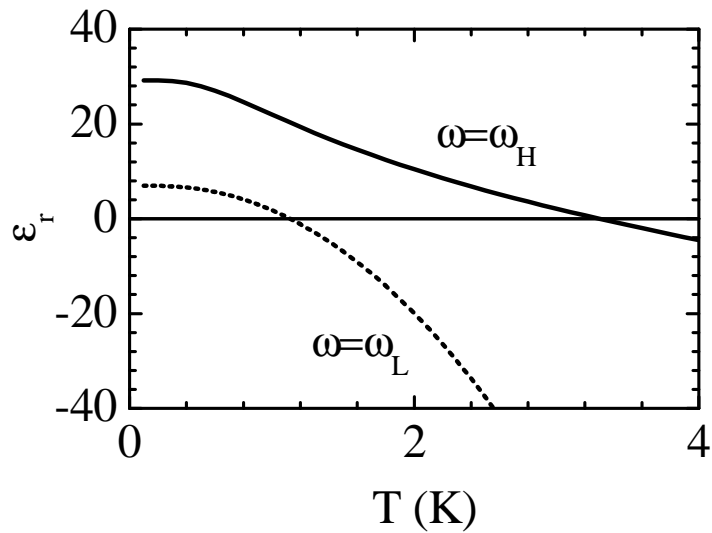,width=3.65in}}
\vspace*{8pt}
\caption{Temperature dependence of the real part of the dielectric
function at the edges of the intra- and intersubband single-particle
continuum $\omega_L$ (dotted curve) and $\omega_H$ (solid curve) at
$q=0.05\times 10^5$ cm$^{-1}$.
}
\label{fig:et}
\end{figure}

At zero temperature or for $T\ll\Delta_{\rm so}$, the intraband
transition is negligible and $\epsilon_r>0$. There is no plasmon
mode in the system. With an increase of the temperature, holes
appear in the valence band and electrons in the conduction band.
The intraband transitions are enhanced and contribute to the
electron-hole propagator of Eq.~(\ref{propagator}) and a dip in
$\epsilon_r$ at the intra-band EHC edge $\omega_L$. This dip in
$\epsilon_r$ results in plasmon modes above $\omega_L$. For
$\Delta_{\rm so}=0$ where $\omega_H=\omega_L$, the intraband
(interband) single-particle continuum occupies the lower (upper)
part of the $\omega-q$ space below $\omega_L$ (above $\omega_L$)
and the plasmon mode are Landau damped. In the presence of the SOI,
i.e. for $\Delta_{\rm so}\neq0$, a gap of width $\omega_H-\omega_L$
is opened between the intra- and interband single-particle continuum
and an undamped plasmon can exist in this gap as shown in
Fig.~\ref{fig:plasmon}. This plasmon mode may perhaps
be observed in experiments.

The appearance of the undamped plasmon mode in the presence
of the SOI is a result of the interplay between the intra- and
the inter-band correlations which can be adjusted by varying the
temperature of the system in experiments. To show the temperature
range in which an undamped plasmon mode exists, in Fig.~\ref{fig:et}
we plot $\epsilon_r (\omega_L)$ (dotted curve) and  $\epsilon_r(
\omega_H)$ (solid curve) as functions of the temperature $T$ at
$q=0.05\times 10^5$ cm$^{-1}$. For $\Delta_{\rm so}=0.08$ meV, an
increase of the temperature from $T=0$ leads to an increase of
the ratio of the intra- to the inter-band correlation while
$\epsilon_r$ in the EHC gap ($\omega_L \leqslant \omega \leqslant
\omega_H$) decreases and crosses zero. There is no undamped plasmon
mode when the inter-band correlation dominates at $T \leq 1.1$ K
and when the intra-band correlation dominates at $T \geq 3.3$ K.
In the temperature regime 1.1 K $\leq T \leq 3.3$ K or $T\approx
2\Delta_{\rm so}$ when the intra- and inter-band correlations
match, however, $\epsilon_r (\omega_L)<0$ while $\epsilon_r
(\omega_H)>0$ and one undamped plasmon mode exists.

In summary: calculating the dynamic dielectric function taking
into account the intrinsic spin-orbit interaction in graphene,
we have studied the collective excitations in graphene. The Dirac
electronic system in graphene is converted into a narrow gap
semiconductor with chiral property by the spin-orbit interaction.
As a result, an undamped collective excitation was found to exist
in the spectral gap of the single-particle continuum and is perhaps
observable in the experiments. More detailed results can be found
elsewhere\cite{wang06}. There have been a steady flow of reports
in the literature on the electronic properties of graphene.
Interestingly, our SOI-dependent dielectric function has recently
been employed to explore the possibility of Wigner crystallization
in graphene\cite{daha}.

\section*{Acknowledgements}

This work has been supported by the Canada Research Chair
Program and a Canadian Foundation for Innovation (CFI) Grant.


\section*{References}

\begin{thebibliography}{0}
\bibitem{graphene_2D}
K.S. Novoselov, et al., PNAS {\bf 102}, 10451 (2005); Science {\bf 306},
666 (2004); Y. Zhang, et al., Phys. Rev. Lett. {\bf 94}, 176803 (2005);
C. Berger, et al., J. Phys. Chem. B {\bf 108}, 19912 (2004).
\bibitem{wallace}
P.R. Wallace, Phys. Rev. {\bf 71}, 622 (1947).
\bibitem{ando}
T. Ando, in {\it Nano-Physics {\&} Bio-Electronics: A New Odyssey},
edited by T. Chakraborty, F. Peeters, and U. Sivan (Elsevier,
Amsterdam, 2002), Chap. 1.
\bibitem{mcclure}
J.W. McClure, Phys. Rev. {\bf 104}, 666 (1956); R.R. Haering and
P.R. Wallace, J. Phys. Chem. Solids {\bf 3}, 253 (1957).
\bibitem{landau_levels}
Y. Zheng and T. Ando, Phys. Rev. B {\bf 65}, 245420 (2002).
\bibitem{qhe_expt}
K.S. Novoselov, et al., Nature {\bf 438}, 197 (2005); Y. Zhang,
Y.-W. Tan, H.L. St\"ormer, and P. Kim, {\it ibid.} {\bf 438},
201 (2005).
\bibitem{high_field}
Y. Zhang, et al., Phys. Rev. Lett. {\bf 96}, 136806 (2006).
\bibitem{qhe_theory}
V.P. Gusynin and S.G. Sharapov, Phys. Rev. Lett. {\bf 95}, 146801
(2005); E. McCann and V.I. Fal'ko, {\it ibid.} {\bf 96}, 086805
(2006).
\bibitem{phys_today}
M. Wilson, Phys. Today {\bf 59} (1), 21 (2006).
\bibitem{chiral}
F.D.M. Haldane, Phys. Rev. Lett. {\bf 61}, 2015 (1988).
\bibitem{kane05} C.L. Kane and E.J. Mele, Phys. Rev. Lett. {\bf 95},
226801 (2005).
\bibitem{wang06} X.F. Wang and T. Chakraborty,  cond-mat/0605498.
\bibitem{sini} N.A. Sinitsyn, et al., Phys. Rev. Lett. {\bf
97}, 106804 (2006).
\bibitem{haldane} F.D.M. Haldane, Phys. Rev. Lett. {\bf 51}, 605 (1983);
F.D.M. Haldane and E.H. Rezayi, {\it ibid.} {\bf 54}, 237 (1985).
\bibitem{nomura}
K. Nomura and A.H. MacDonald, Phys. Rev. Lett. {\bf 96}, 256602
(2006).
\bibitem{goerbig06} M.O. Goerbig, R. Moessner, and B. Doucot,
cond-mat/0604554.
\bibitem{fisher06} J. Alicea and M.P.A. Fisher, Phys. Rev. B
{\bf 74}, 075422 (2006).
\bibitem{fano} G. Fano, F. Ortolani, and E. Colombo, Phys. Rev. B {\bf 34},
2670 (1986).
\bibitem{laughlin} R.B. Laughlin, Phys. Rev. Lett. {\bf 50}, 1395 (1983).
\bibitem{fqhe_book} T. Chakraborty and P. Pietil\"ainen, {\it The
Quantum Hall Effects} (Springer, Heidelberg, 1995), 2nd edition.
\bibitem{gr-fqhe} V.M. Apalkov and T. Chakraborty, Phys. Rev. Lett.
{\bf 97}, 126801 (2006).
\bibitem{fqhe_review} T. Chakraborty, Adv. Phys. {\bf 49}, 959 (2000).
\bibitem{wojs} A. Wojs and J.J. Quinn, Phys. Rev. B {\bf 66}, 45323
  (2002).
\bibitem{vint}B. Vinter, {\it Phys. Rev.} {\bf B15}, 3947 (1977).
\bibitem{wang1}X.F. Wang, {\it Phys. Rev.} {\bf B72}, 85317 (2005).
\bibitem{pere}N.M.R. Peres, F. Guinea, and A.H. Castro Neto,
Phys. Rev. {\bf 72}, 174406 (2005); J. Nilsson, A.H. Castro
Neto, N.M.R. Peres, and F. Guinea, cond-mat/0512360 (2005).
\bibitem{mele} D.P. DiVincenzo and E.J. Mele, {\it Phys. Rev.} {\bf B29},
1685 (1984).
\bibitem{daha}H.P. Dahal, Y.N. Jogelkar, K.S. Bedell, and A.V. Balatsky,
cond-mat/0609440 (2006).

\end{thebibliography}
\end{document}